\documentclass[twocolumn]{aastex631}

\usepackage[utf8]{inputenc}
\usepackage{graphicx}
\usepackage{indentfirst}
\usepackage{amsmath}
\usepackage{natbib}
\setcitestyle{aysep={}}
\usepackage{rotating}

\newcommand{\msun}{$\mathrm{M}_{\odot}$}

\newcommand{\ha}{H$\alpha$}

\begin{document}

\title{Star-Forming Nuclear Clusters in Dwarf Galaxies Mimicking AGN Signatures in the Mid-Infrared}

\author[0000-0003-1055-1888]{Megan R. Sturm}
\affil{Department of Physics, Montana State University, Bozeman, MT 59717, USA}

\author[0009-0009-6661-6194]{Bayli Hayes}
\affil{Department of Physics, Montana State University, Bozeman, MT 59717, USA}

\author[0000-0001-7158-614X]{Amy E. Reines}
\affil{Department of Physics, Montana State University, Bozeman, MT 59717, USA}


\begin{abstract}
    
Effectively finding and identifying active galactic nuclei (AGNs) in dwarf galaxies is an important step in studying black hole formation and evolution. In this work, we examine four mid-IR-selected AGN candidates in dwarf galaxies with stellar masses between $M_\star \sim 10^8 - 10^9$ \msun, and find that the galaxies are host to nuclear star clusters (NSCs) that are notably rare in how young and massive they are. We perform photometric measurements on the central star clusters in our target galaxies galaxies using {\it Hubble Space Telescope} optical and near-IR imaging and compare their observed properties to models of stellar population evolution. We find that these galaxies are host to very massive ($\sim10^7 M_\odot$), extremely young ($\lesssim 8$ Myr), dusty ($0.6 \lesssim \mathrm{A_v} \lesssim 1.8$) nuclear star clusters. Our results indicate that these galactic nuclei have ongoing star-formation, are still at least partially obscured by clouds of gas and dust, and are most likely producing the extremely red AGN-like mid-IR colors. Moreover, prior work has shown that these galaxies do not exhibit X-ray or optical AGN signatures. Therefore, we recommend caution when using mid-IR color-color diagnostics for AGN selection in dwarf galaxies, since, as directly exemplified in this sample, they can be contaminated by massive star clusters with ongoing star formation.

\end{abstract}



\section{Introduction}
\label{sec:Introduction}

Massive black holes (BHs) with masses of $M_{\rm BH} \sim 10^6-10^9$ \msun\ are highly prevalent objects in the Universe, living in the centers of almost all massive ($\gtrsim 10^{10}$ \msun) galaxies \citep[e.g.,][]{Magorrian1998}. Multiple seed formation scenarios have been proposed \citep{Bond1984,Loeb1994, Begelman2006}, however we do not have enough observational evidence yet to identify which gave rise to the population of massive BHs that we observe in the early and present-day Universe. Although the {James Webb Space Telescope (JWST)} has pushed the boundaries of observable active galactic nuceli (AGN) in the Universe to new limits \citep{Onoue2023, Kocevski2023, Maiolino2023a}, the origins and earliest stages of growth for these BH seeds are still out of direct observational reach \citep{Volonteri2016,schleicher2018, Vito2018}. 

Since high-redshift seed BHs remain observationally elusive, as a proxy for directly observing the initial formation and growth of such BHs, we can instead study massive BHs with $M_{\rm BH} \lesssim 10^5$ \msun\ residing in dwarf galaxies ($M_{\star} \lesssim 10^{9.5}$ \msun) in the local Universe. Following from the idea of hierarchical assembly, dwarf galaxies tend to have undergone fewer mergers in comparison to more massive galaxies keeping their BHs from growing as much \citep{volonteri2008} and, indeed, they host the least-massive BHs known from an observational standpoint \citep[for a review, see][]{Reines2022}. Therefore, identifying and characterizing BHs in local dwarf galaxies can place observational constraints on seed BHs from the early Universe.

One method for identifying massive BHs is through observations of AGNs, which emit radiation across the full electromagnetic spectrum. Various diagnostic diagrams \citep[e.g., the BPT optical emission line diagram;][]{Baldwin1981} can distinguish emission originating from AGNs versus star-formation processes. In the mid-infrared (mid-IR), the characteristic power-law spectrum of dusty AGNs can be used as a diagnostic and has the advantage of suffering minimally from nuclear and galaxy scale obscuration. Multiple works have proposed mid-IR color-color diagnostics to distinguish between emission originating from dust heated by an AGN and dust heated by stellar processes, which will typically be at a much lower temperature \citep{Jarrett2011, Mateos2012, Stern2012}. Combining these diagnostics with observations from the all-sky mid-IR {\it Wide-field Infrared Survey Explorer} \citep[WISE,][]{Wright2010} has produced large samples of AGNs and quasars \citep{Secrest2015, Assef2018} with high reliability in moderate to high-mass galaxies.

\begin{figure}
    \centering
    \includegraphics[width=3in]{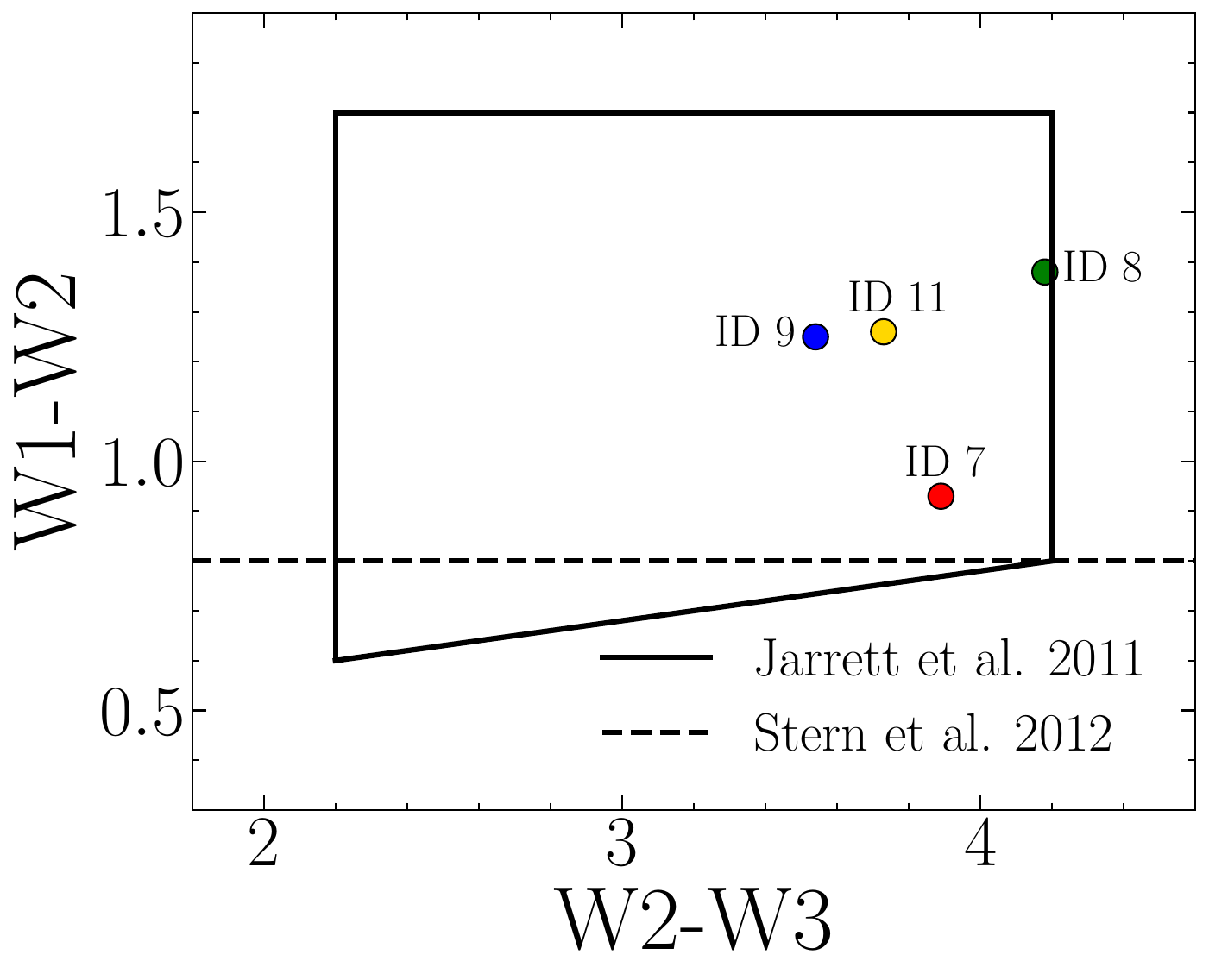}
    \caption{WISE color-color diagram for the four BPT star-forming dwarf galaxies in our sample. We show the AGN selection box from \citet{Jarrett2011} as the solid line and the W1-W2$>$0.8 cut from \citet{Stern2012} as the dashed line. Following both of these diagnostics, our galaxies are categorized as AGNs despite a lack of evidence at optical and X-ray wavelengths \citep{Latimer2021}.}
    \label{fig:wise}
\end{figure}

\begin{figure*}
    \centering
    \includegraphics[width=7in]{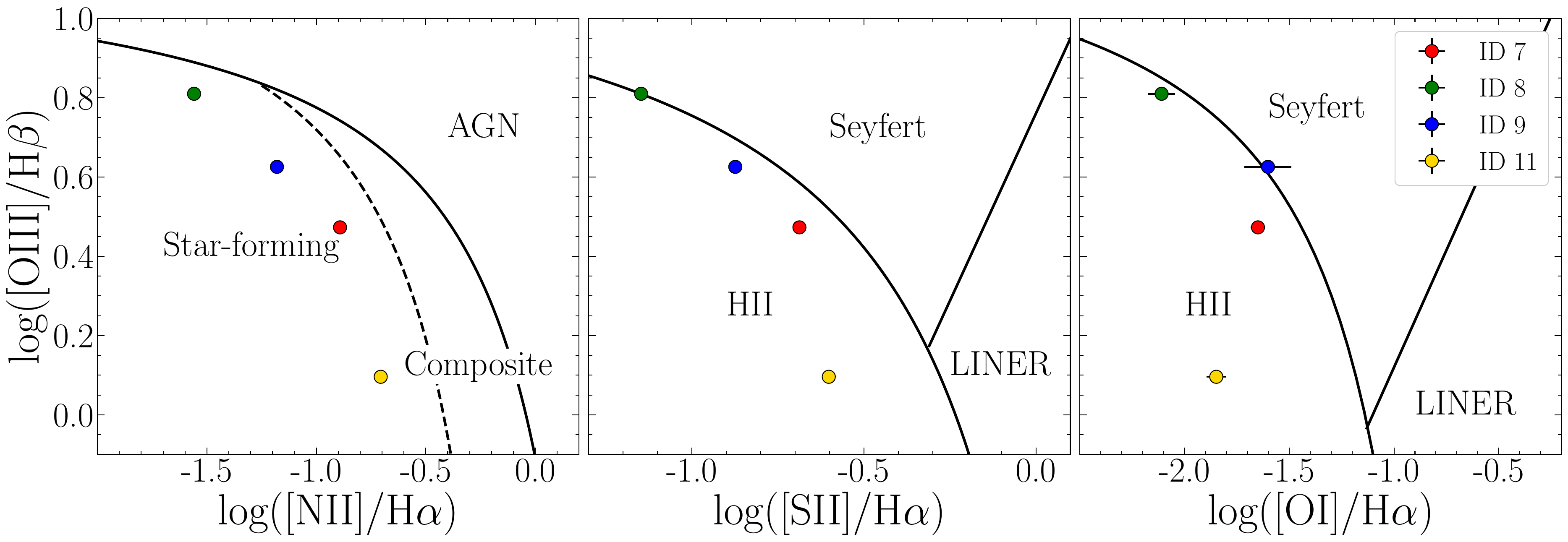}
    \caption{Narrow emission line ratio diagnostic diagrams for the four galaxies in our sample. Values for our sample come from \citet{Latimer2021} using the methods described in \citet{Reines2013}. The lines separating different regions in each plot are from \citet{Kewley2006}. Our galaxies are consistent with being dominated by star-formation based on all of these diagnostic diagrams.}
    \label{fig:bpt}
\end{figure*}

However, multiple works have attempted to extrapolate these mid-IR diagnostics to the dwarf galaxy regime and find that the AGN fraction increases at low mass \citep{Satyapal2014, Sartori2015}, contradictory to findings at other wavelengths. {\it WISE} has a relatively low resolution ($\sim 6"$), which both complicates the cross-matching process to identify the correct host galaxies \citep{lupi2020} and means that, in addition to emission from a potential AGN, emission from the host galaxy will also be present. At low mass, AGNs have lower Eddington limits, meaning they must be highly accreting in order to be observed over the stellar/star-formation-related emission from the host galaxy. Even looking at a sample of highly-accreting, optically-selected AGNs \citep{Reines2013}, \citet{Hainline2016} find that the majority are {\it not} selected as AGNs in WISE since the mid-IR emission is dominated by the host galaxy.

It is generally rare for BHs to be accreting at their Eddington limit \citep{Schulze2010}, however, as seen at higher galaxy mass, selecting galaxies based on AGN-like WISE colors does create a bias towards bolometrically dominant AGNs. Moreover, it is possible for a $\sim10^4$ \msun \: BH to have bolometric luminosity equivalent to that of a $\sim10^9$ \msun \: dwarf galaxy ($\sim10^{42}$ erg s$^{-1}$). 
However, if these AGNs are highly accreting, then it is notable that we do not always observe complementary optical or X-ray AGN signatures \citep{Latimer2021}. On the other hand, many AGNs are Compton-thick and about half of mid-IR selected AGNs are heavily obscured \citep[e.g., Fig. 1 in][]{Petter2023}. Additionally, photoionization modeling has shown that at low metallicity/BH mass, BPT diagnostics likely fail \citep{Groves2006, Cann2019}, and the literature appears to support X-ray emission in dwarf galaxy AGNs being lower than in their more massive counterparts \citep[e.g.,][]{Simmonds2016}, including even in BPT AGNs with broad lines \citep{Baldassare2017} and BPT AGNs with red WISE colors \citep{Latimer2021}. In any case, it is important to consider other explanations for the very red mid-IR colors of dwarf galaxies, particularly for those without other supporting evidence for AGNs.

Dwarf starburst galaxies have been observed to heat dust enough to produce very red mid-IR colors \citep{Griffith2011, Izotov2011}. In this case, invoking the presence of an AGN is not even necessary to produce the extreme mid-IR colors. Notably, \citet{Hainline2016} find that dwarf starburst galaxies can mimic the mid-IR colors of AGNs. They demonstrate that a single $W1-W2$ AGN color selection is subject to severe contamination from dwarf starburst galaxies, finding that dwarf galaxies with the reddest mid-IR colors also have the youngest stellar populations and highest star formation rates.

In a follow-up study, \citet{Latimer2021} present {\it Chandra X-ray Observatory} and {\it Hubble Space Telescope (HST)} observations of a subset of dwarf galaxies from the \citet{Hainline2016} sample that fall in the \citet{Jarrett2011} mid-IR two-color AGN selection box. Roughly half of the galaxies have optical emission line ratios indicating an AGN, while the remaining galaxies are classified as star-forming based on the optical BPT diagram. While nearly all of the optically-selected AGNs have detectable X-ray point sources with luminosities exceeding that expected from star formation, \citet{Latimer2021} do not find compelling evidence for AGNs in the star-forming dwarf galaxies based on their X-ray analysis.

In this work, we seek to determine the stellar properties of the BPT star-forming galaxies that have mid-IR colors mimicking AGNs using observations from {\it HST}. In Section \ref{sec:Sample Selection} we describe our sample of star-forming dwarf galaxies. In Section \ref{sec: Optical Photometry} we describe the observations and our method of performing photometry on the {\it HST} optical images. Finally, we compare colors and spectral energy distributions of the central star clusters with models of stellar cluster evolution to estimate cluster ages, extinctions, and masses in Section \ref{sec: Modelling Star Cluster Evolution}. A discussion and our conclusions are presented in Section \ref{sec: Conclusions}.

\begin{centering}
\begin{table*}

\centering

\begin{tabular}{cccccccccc}

\toprule
ID & SDSS Name & NSAID & RA & DEC & z & log (M$_*$/M$_\odot$) & F140W & F336W & F606W \\\hline

7 & J005904.10+010004.2 & 6205 & 00:59:04 & 01:00:04 & 0.01743 & 8.7 & 21.24 & 19.17 & 19.61 \\
8 & J154748.99+220303.2 & 98135 & 15:47:49 & 22:03:03 & 0.03154 & 8.2 & 21.77 & 18.77 & 19.19 \\
9 & J160135.95+311353.7 & 57649 & 16:01:36 & 31:13:54 & 0.03085 & 8.6 & 21.90 & 20.28 & 20.32 \\
11 & J233244.60$-$005847.9 & 151888 & 23:32:45 & -00:58:46 & 0.02437 & 9.3 & 20.95 & 19.39 & 19.62 \\

\hline     
\end{tabular}

\caption{Column 1: Galaxy ID from \citet{Latimer2021}. Column 2: SDSS name. Column 3: NSAID. Column 4: Right ascension in units of hours:minutes:seconds. Column 5: Declination in units of degrees:arcminutes:arcseconds. Column 6: redshift from NSA parameter zdist. Column 7: log galaxy stellar mass from \citet{Latimer2021}. Columns 8-10: ST magnitudes of the central star clusters in the F140W, F606W and F336W filters.}
\label{tab:sample_phot}

\end{table*} 
\end{centering}

\begin{figure*}
    \centering
    \includegraphics[width=18cm]{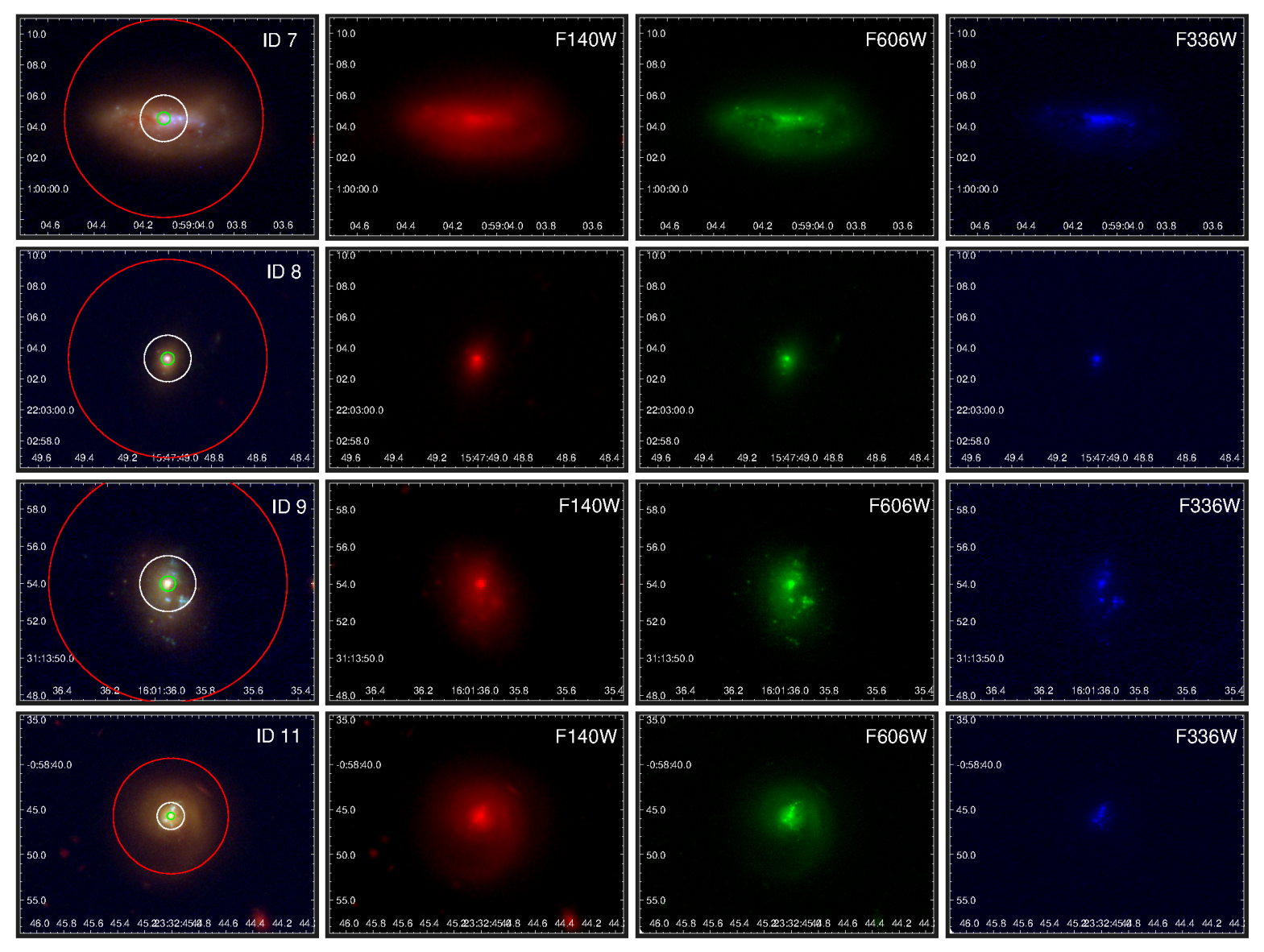}
    \caption{Three-color {\it HST} images of the four BPT star-forming galaxies in our sample. The red images represent the near-infrared (NIR) band (F140W filter), green images show the optical band (F606W filter) and blue images show U/UV band (F336W filter). In the left column, the green circles show the peak of the NIR emission in the {\it HST} images with radius of 0\farcs4, which we define as our central star clusters, the red circles show the WISE W2-band resolution of 6\farcs4 and the white circles show the SDSS 3\farcs0 spectroscopic apertures.
    }
    \label{fig:HST_threecolor}
\end{figure*}


\section{Sample of Star-forming Galaxies}
\label{sec:Sample Selection}

Our sample consists of four dwarf galaxies (listed in Table \ref{tab:sample_phot}) that are classified as hosting AGNs using {\it WISE} mid-IR selection techniques \citep{Hainline2016}, but they do not show optical or X-ray AGN signatures \citep{Latimer2021}. These galaxies were originally part of the \citet{Hainline2016} sample of $\sim$18,000 dwarfs (M$_* \lesssim 3 \times 10^9 \ \mathrm{M}_\odot$) in the NASA-Sloan Atlas (NSA) with significant mid-IR detections in the first three bands of the ALLWISE data release. \citet{Hainline2016} find 41 of these galaxies having mid-IR colors consistent with an AGN; having $W1-W2$ versus $W2-W3$ colors falling within the \citet{Jarrett2011} {\it WISE} AGN selection box (shown in Figure \ref{fig:wise}).

In a follow-up X-ray and optical study, \citet{Latimer2021} selects 11 of these 41 galaxies that have both optical emission-line measurements (redshifts and fluxes) and signal-to-noise ratios $>$ 5 in all four {\it WISE} bands. Following \citet{Reines2013} methods and using SDSS spectra, they find that, despite the fact that all 11 of the galaxies in their sample are considered mid-IR AGNs, five lie in the star-forming region of the BPT diagram according to the classification scheme of \citet[][see Figure \ref{fig:bpt}]{Kewley2006}. For this work, we select the four galaxies from these five BPT star-forming galaxies that have {\it HST} imaging to study further. 

Three of our galaxies (IDs 7, 8 and 9) were not detected at all in the {\it Chandra} X-ray imaging presented in \citet{Latimer2021}, placing an upper limit on their X-ray luminosities of $\mathrm{L_{2-10 keV}} \lesssim 10^{39.6}$ erg s$^{-1}$. All three of these upper limits are consistent with the expected contributions from high-mass X-ray binaries (XRBs). The last target (ID 11) does have an X-ray point source detected with $\mathrm{L_{2-10 keV}} = 39.8$ erg s$^{-1}$. However, this X-ray point source is offset from the center of the galaxy and the brightest near-IR source, and the luminosity is consistent with the expected contribution from XRBs or an ultra-luminous X-ray source (ULX). Interestingly, in addition to its AGN-like mid-IR colors, ID 11 was found to also show variability in the mid-IR indicative of the presence of an AGN\footnote{We check multiple other works for variability signatures in our other galaxies but did not find any \citep{Baldassare2018, Baldassare2020, Burke2022, Wasleske2022, Aravindan2024}.} \citep{Secrest2020}. Despite having mid-IR colors (and, in the case of ID 11, variability in the mid-IR) that look AGN-like, there is not compelling evidence at optical or X-ray wavelengths that these dwarf star-forming galaxies host AGNs.


\section{{\it HST} Optical and Near-IR Photometry}
\label{sec: Optical Photometry}

\subsection{Observations}
\label{sec:observations}

The {\it HST} observations for these four galaxies were taken in 2019 (Proposal 15607, PI: Reines; found at \dataset[10.17909/rdw9-n374]{http://dx.doi.org/10.17909/rdw9-n374}) and were first presented in \citet{Latimer2021}. We use HST/WFC3 UVIS and IR imaging in three different filters: F336W ($\sim$ U-band), F606W ($\sim$ wide V-band) and F140W (JH gap). During these observations, each galaxy was observed for one orbit with exposure times of $\sim$7-9 minutes in the IR F140W filter, $\sim$11-12 minutes in the UVIS F606W filter, and $\sim$12-15 minutes in the UVIS F336W filter. The images were processed using the automated AstroDrizzle pipeline \citep{Gonzaga2012}. 

We make relative astrometry corrections by manually aligning stars in the F336W and F140W filters to the corresponding stars in the F606W filter and adjusting the WCS for the first two accordingly. This resulted in astrometric shifts up to 0\farcs16. This step ensures that the central star clusters of interest lie directly on top of each other in all filters. Three-color {\it HST} images of our galaxies are shown in Figure \ref{fig:HST_threecolor}.

\subsection{Aperture Photometry Measurements}
\label{sec:opt_phot}

We are interested in the photometric properties of the central star cluster in each galaxy. The center point of the cluster is selected as the peak of the NIR emission in the {\it HST} images following \citet{Latimer2021}. Our star clusters generally reside within a radius of $\sim$0\farcs4, so we use this aperture size in the photometry measurements for all four of our targets for consistency. 

We find source counts by performing aperture photometry in each filter using the \texttt{photutils} software package \citep{Bradley2017}. We estimate the background counts within the target aperture by first finding the median background counts per area within an annulus, surrounding the target aperture. The total source counts is then calculated as the counts within the circular aperture ($r_\mathrm{aperture}=0\farcs4$) minus the median counts per area from the background annulus times the area of the circular aperture. The dominant source of uncertainty in our measurements comes from our choice of the background annulus, so we vary both the inner radius of the annulus (from 0\farcs5 to 0\farcs8) and the width of the annulus (from 0\farcs1 to 0\farcs3) and find errors $\lesssim$15\% for all targets/filters. 

Lastly, we perform aperture corrections based on our target aperture size of 0\farcs4 (0.84 for the F140W filter, 0.91 for the F606W filter and 0.89 for the F336W filter). We report our measured ST magnitudes in each filter in Table \ref{tab:sample_phot}.


\section{Properties of the Star Clusters}
\label{sec: Modelling Star Cluster Evolution}

We compare our broadband photometric measurements to a GALEV simple stellar population (SSP) synthesis model \citep{Kotulla2009} in order to estimate the ages, extinctions, and masses of the central star clusters in our target galaxies. We opt for using GALEV over other models because it provides the option of including a set of metallicity-dependent emission lines \citep{Anders2003}, which make up a  non-negligible portion of the flux at optical wavelengths in very young stellar clusters \citep{Reines2010}. Our GALEV model uses the Padova isochrones, a timestep of 4 Myr, a Kroupa initial mass function (IMF) (0.1-100\msun), a metallicity of [Fe/H] = -0.7 (as appropriate for these galaxies; \citealt{Latimer2021}) and an initial mass of 10$^5$ \msun. 

Uncertainties in derived quantities using stellar population synthesis models arise from theoretical/observational unknowns, including those associated with different phases of stellar evolution and the IMF. Additionally, there is an age-metallicity degeneracy, where an older, metal-poor population cannot always be distinguished from a younger, metal-rich population \citep{Worthey1994}. However, \citet{Conroy2009} find that including these model uncertainties does not strongly impact physical properties such as star formation rates, stellar masses, ages, and metallicities. Specifically, they state that stellar masses have errors of $\sim$0.3 dex. Furthermore, looking at their GALEV model $V-K$ colors versus time, \citet{Kotulla2009} find that the color evolution agrees with true colors/ages for star clusters in the Large Magellanic Cloud within 0.4 dex.

While there are inherent uncertainties associated with using SSP models, our choice of filters helps mitigate the effects of these uncertainties in drawing conclusions about the nature of the observed star clusters. We have photometric observations in three filters, including both red and blue optical filters and a NIR filter. When comparing observed SEDs to SSP models, this variety of filters provides a balanced, representative distribution in age, metallicity and extinction, as demonstrated in \citet{Degrijs2003}. Specifically, our choice of filters includes both the U and V bands, which are found to be important in determining cluster ages \citep{Anders2004}. Additionally, \citet{Degrijs2003} show that assuming a generic, subsolar metallicity, as we do in this work, results in increased scatter in the age distribution of clusters, however, the peak generally remains the same.

\begin{figure}
\centering
\includegraphics[width=3.3in]{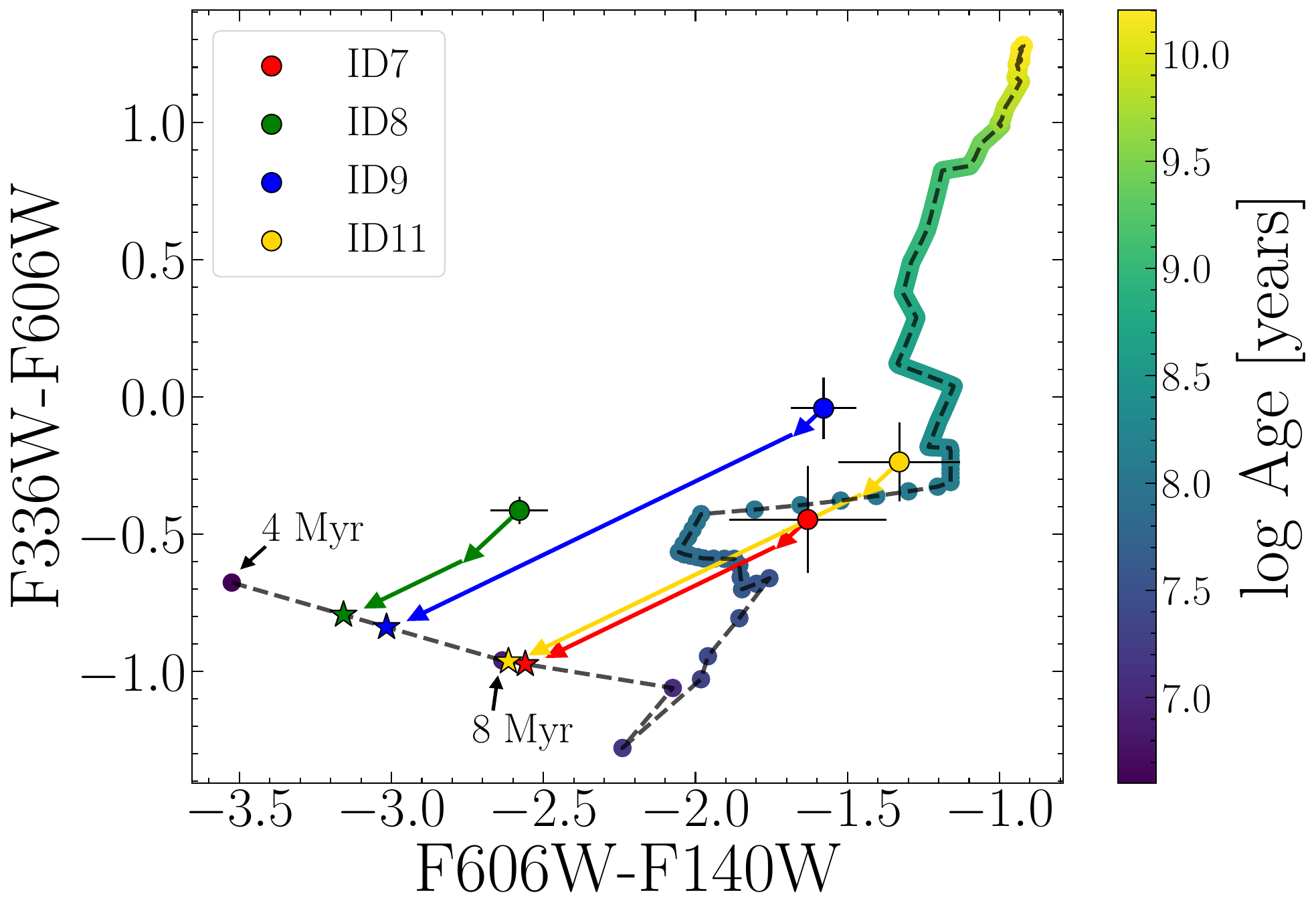}
\caption{F336W-F606W versus F606W-F140W color-color diagram. We plot the expected color-color evolutionary track from the GALEV models (color-coded by age). We also plot the observed colors for each galaxy as circles. We draw de-reddening vectors from each point, accounting for both extinction within the Milky Way and internal extinction within the target galaxy. The intersection point for each of the de-reddening vectors with the model is shown with a star. All of the clusters intersect with the GALEV model at ages $\lesssim$ 10 Myr, confirming that these are indeed young stellar clusters with significant amounts of dust extinction. See \S\ref{sec:ages}. Error bars represent uncertainties resulting from the photometric measurements.}
\label{fig:color_color}
\end{figure}

\subsection{Age and Extinction Estimates}
\label{sec:ages}

We estimate the ages of our stellar clusters using a color-color diagram. We plot the observed F336W-F606W versus F606W-F140W  color for each of our targets along with the expected trajectory from the GALEV model in Figure \ref{fig:color_color}. However, we expect these observed colors to be reddened by dust both in the target galaxies themselves as well as within the Milky Way. We account for Galactic extinction within the Milky Way using the \citet{Cardelli1989} extinction curve with the typical value of R$_{\mathrm{v}}$=3.1 and A$_\mathrm{v,Gal}$ values corresponding to the location of each galaxy from \citet{schlegel1998}, ranging from A$_\mathrm{v,Gal}$=0.08-0.2. Given the mid-IR colors of our galaxies, we expect significant internal extinction within the target galaxies. Following the work of \citet{Reines2008}, we account for the internal extinction using a 30 Doradus extinction curve \citep{fitzpatrick1999} and using a R$_\mathrm{v}$ value of 4.48 from \citet{demarchi2014}. \footnote{We use R$_\mathrm{v}$=4.48 to more accurately model the dusty environment expected within the target galaxies, however, we note that using the typical Milky Way value of 3.1 only increases our age estimates by 0.1 dex (1-2 Myr) and does not change our cluster mass estimates at all.} This allows us to get a direction for the internal reddening vectors in F336W-F606W vs.\ F606W-F140W color space.

We extend de-reddening vectors for each cluster to find where their inferred intrinsic colors intersect with our GALEV models (shown in Figure \ref{fig:color_color}). Since these clusters are presumably responsible for the red mid-IR colors in {\it WISE}, we expect them to be young with significant dust extinction at the shorter wavelengths explored here. Therefore, while the clusters in ID 7 and ID 11 intersect the models at both young and old ages (once at $\sim10^7$ years, again at $\sim10^{7.8}$ years and lastly $\sim10^8$ years), we think it is much more likely that they are young with A$_\mathrm{v} > 1$. Under this assumption, all four clusters intersect with the GALEV model color-color evolutionary track within $\sim$the first 2 points (ages $\lesssim$ 10 Myr). We interpolate between the first three points in the model to find a more precise age for the clusters ($\sim$ 5 to 8 Myr) and report these values in Table \ref{tab:agemass}. 

As a consistency check, we make an additional age measurement for our star clusters using the equivalent width of the H$\alpha$ emission line, which quantifies the ratio of ionizing radiation from star formation to the continuum flux density as follows: 

\begin{equation}
\begin{gathered}
    \mathrm{W}_{\mathrm{H}\alpha} = \frac{\mathrm{F}_{\mathrm{H}\alpha}}{f_{\mathrm{cont}}} 
\end{gathered}
\end{equation}

\noindent
where ${\mathrm{F}_{\mathrm{H}\alpha}}$ is the flux of the H$\alpha$ emission line and $f_{\mathrm{cont}}$ is the underlying continuum flux density.

We use H$\alpha$ equivalent width values for our targets from the NSA and compare the values to those predicted by the Starburst99 model \citep{Leitherer1999}. We use the Starburst99 model with instantaneous star formation, IMF $\alpha$=2.35, Z=0.004 (corresponding to the metallicity of our GALEV model) and M$_\mathrm{up}$=100 \msun. These ages are also reported in Table \ref{tab:agemass}. We calculate nearly identical ages using the GALEV color-color model and the Starburst99 H$\alpha$ equivalent width. 

Since we find remarkably similar age measurements using a completely different method from a separate model, we consider our age estimates to be robust. Furthermore, we consider our SSP-derived ages robust since, in their work, \citet{Kocharov2018} perform composite stellar population fits with detailed star formation histories and find that the light-weighted age measurements using these fits are in good agreement with those calculated just using SSPs. Additionally, \citet{Walcher2006} compare ages calculated by fitting SEDs to composite versus SSP models. They find that for the nuclei with older stellar populations, the $\chi^2$ value for the composite fits are significantly better, however, for their one galaxy with significant populations of young stars and age $\sim10^7$years, similar to the ages we find for our sample, the $\chi^2$ value for the composite versus SSP models are comparable.

We calculate the expected internal exitnction from our color-color diagram and find A$_\mathrm{v}$ values in the range $0.6 \lesssim \mathrm{A_v} \lesssim 1.8$. We also find that the nebular extinction in these galaxies, calculated from the Balmer decrement and using \ha \: and H$\beta$ fluxes as reported in the NSA, results in A$_\mathrm{v,nebular}=0.36-0.63$. These are smaller than the stellar A$_\mathrm{v}$ calculated from our color-color diagram, which is likely due to the fact that our photometry probes just the central star clusters (with aperture radius of 0\farcs4), while the NSA fluxes come from SDSS spectroscopy (with fiber diameter of 3\farcs0). The NSA fluxes come from a much larger region that includes more diffuse areas in the outer regions of the galaxies that are expected to have lower extinction.

\subsection{Star Cluster Mass Estimates}
\label{sec:masses}

\begin{figure*}
\centering
\includegraphics[width=7in]{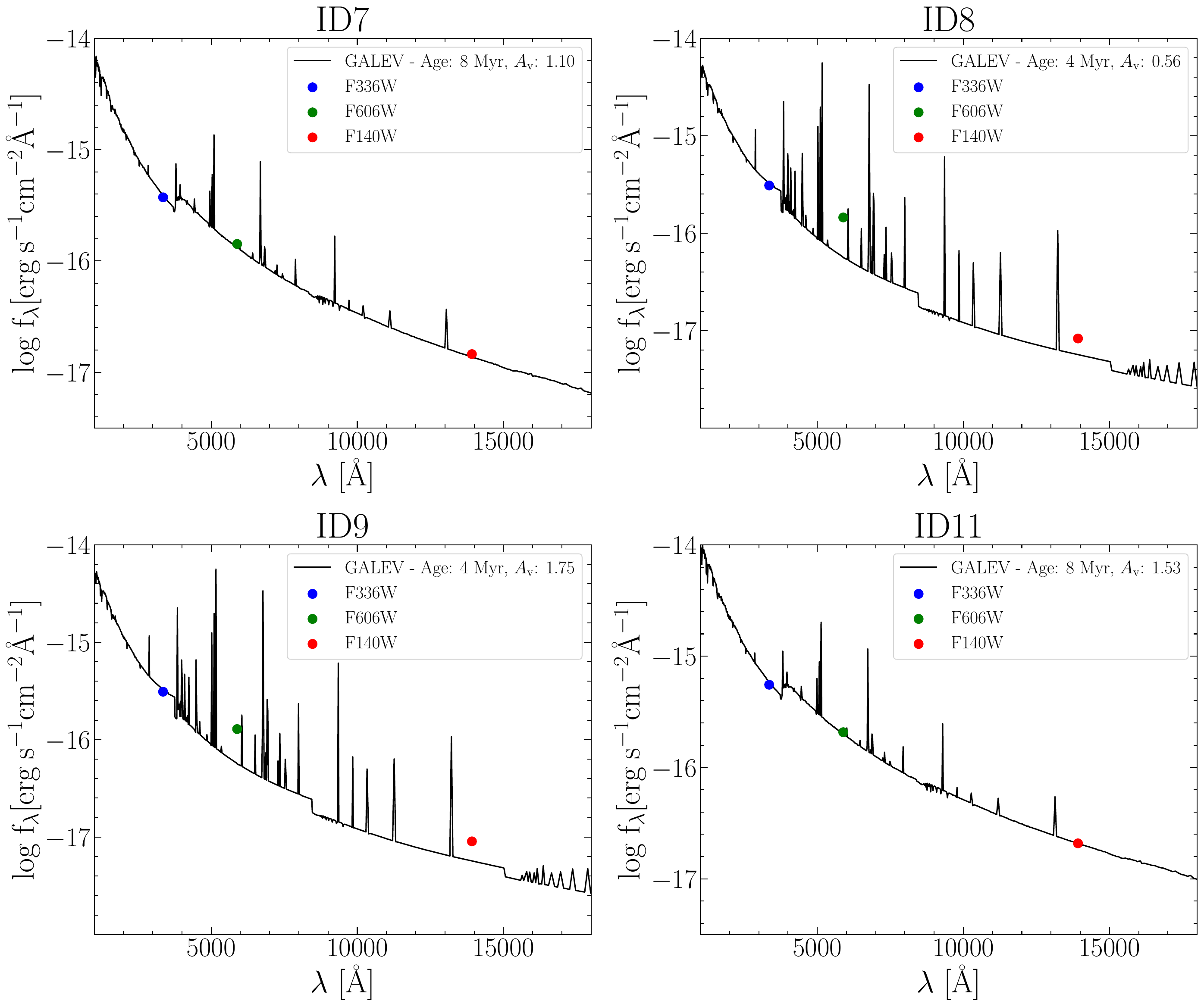}
\caption{GALEV model spectra corresponding to the color-color-derived age of each galaxy. For each of these spectra we scale them to match the observed photometry of our galaxies and apply the expected reddening. We overlay our photometric observations in each filter as colored points.}
\label{fig:spectra}
\end{figure*}

We estimate masses of the star clusters as follows. Using the age and extinction estimates derived from the color-color diagram in Figure \ref{fig:color_color}, we compare our observed photometric measurements to the model spectra at the closest age for our clusters (4 Myr for IDs 8 and 9, and 8 Myr for IDs 7 and 11; see Figure \ref{fig:spectra}). In order to compare our observed photometry measurements to the GALEV model spectra, we de-redden the observed fluxes by removing the extinction estimated from Section \ref{sec:ages}. We then shift our GALEV model spectra to the corresponding redshift of each galaxy and convolve it with the {\it HST} filter throughput curves in order to derive simulated photometry in our {\it HST} bands. Since the GALEV spectra are normalized to a distance of 10 pc, we scale the model flux densities according to the distance of each galaxy.

We compare the model photometry with our de-reddened, observed photometry in each filter to find a scaling between the two. In the models, the flux density is linearly proportional to the mass of the cluster. Therefore, we are able to use this scaling along with the initial model mass of $10^5$\msun (the stellar mass does not change significantly between the first two points in the model) to estimate masses of the observed clusters. We use the scaling that we calculate from the F336W filter, since the F606W filter can be significantly impacted by emission lines \citep{Reines2010} and the F140W filter could be boosted by red supergiants and/or very hot dust\footnote{We note that using the scaling calculated from the F140W filter results in mass estimates that are larger by $\sim$0.1 dex.} \citep{Reines2008}. Our estimated star cluster masses are reported in Table \ref{tab:agemass}. All of our cluster masses are $\sim$10$^{7}$ \msun.

\begin{table}
\centering
\vspace{0.5cm}
\begin{tabular}{cccccccc}
\toprule
   ID & log Age (colors) & log Age (H$\alpha$EW) & log M$_*$ &  A$_\mathrm{v}$  \\
     & [years] & [years] & [M$_\odot$] & [mag] \\
\hline
 7 & 6.9 & 6.9 & 6.9 & 1.10 \\
 8 & 6.8 & 6.7 & 7.0 & 0.56 \\
 9 & 6.8 & 6.8 & 7.0 & 1.75 \\
11 & 6.9 & 6.8 & 7.3 & 1.53 \\
\hline
\end{tabular}
\caption{Column 1: Galaxy ID. Column 2: Log cluster age in years estimated from the color-color diagram in Figure \ref{fig:color_color}.
Column 3: Log cluster age in years estimated by comparing the H$\alpha$ equivalent width from the NSA to the Starburst99 models. Column 4: Log cluster mass in solar masses. Column 5: Internal $A_V$ due to dust extinction. Uncertainties from using stellar population synthesis models are $\sim$0.3 dex for stellar mass \citep{Conroy2009} and $\sim$0.4 dex for age \citep{Kotulla2009}.}
\label{tab:agemass}
\end{table}


\section{Discussion and Conclusions}
\label{sec: Conclusions}

In this work, we investigate the properties of the central star clusters in a sample of four dwarf galaxies that exhibit very red mid-IR colors, leading to them being categorized as AGNs using typical color-color diagnostics for {\it WISE} observations. Contrary to the pro-AGN evidence in the mid-IR, they do not have optical narrow emission line ratios or X-ray signatures supporting the case for AGNs \citep{Latimer2021}. Furthermore, the mid-IR color evidence for AGNs is not all that compelling; our galaxies follow the general trend of optically-selected star-forming dwarf galaxies in the WISE color-color diagram \citep[see Figure 6 in][]{Hainline2016} and may have scattered into the \citet{Jarrett2011} AGN selection box due to relatively hot dust \citep[see Figure 7 in][]{Hainline2016}.
We use {\it HST} optical and near-IR photometry in conjunction with stellar population synthesis models (GALEV and Starburst99) to characterize the properties of the central star clusters in these galaxies. The clusters coincide with the peak of the near-IR emission in each galaxy and we hypothesize that they are responsible for the red mid-IR colors in {\it WISE}.

We derive stellar masses of $\sim 10^7 M_\odot$ for the central star clusters in our sample. These masses are consistent with  nuclear star clusters (NSCs; \citealt{Neumayer2020}) and higher than typical young super star clusters \citep[e.g.,][]{Melo2005, Reines2008} and globular clusters \citep{Gratton2019, Baumgardt2019}. NSCs are most commonly found in galaxies having stellar masses $8 < \mathrm{log \: (M_{stellar}/M_\odot) < 10}$, with $\sim80\%$ of both early and late-type galaxies of mass $\sim10^9$\msun \: hosting an NSC \citep{Boker2002, Carollo2002, denbrok2014, Georgiev2014, Ordenes2018, sanchez2019}. For comparison, the host galaxy stellar masses of our targets are in the range $8.2 < \mathrm{log \: (M_{stellar}/M_\odot) < 9.3}$. The NSC masses are at the high end, but consistent within the scatter, of the linear scaling relation found between host galaxy stellar mass and NSC mass from \citet[][ equation 1 in their paper]{Neumayer2020} found by combining multiple samples of NSCs. 

Furthermore, we find that the NSCs in our sample of dwarf galaxies are likely quite dusty, with internal A$_\mathrm{v}$ values between 0.56 and 1.75. While these A$_\mathrm{v}$ values are not particularly high, we note that they are derived using optical emission, which is not able to penetrate the potential denser regions of clumpy dust surrounding the nucleus which would be emitting in the mid-IR \citep{Gordon1997, Reines2008}. Nonetheless, the A$_\mathrm{v}$ values for these clusters indicate they are still at least partially enshrouded in dust, aligning with the fact that our NSCs are quite young; we find that the ages of our clusters are all $\lesssim$ 8 Myr. 

Our sample of dwarf galaxies are likely host to NSCs that formed (or are in the process of forming) via in situ star formation. In this scenario, a burst of star formation is caused by gas infalling to the central few parsecs of the galaxy, and it is thought to be the dominant growth mechanism for NSCs in late-type galaxies \citep{Neumayer2020}. Concentrations of young stars have also been observed in the NSCs of the Milky Way \citep{Paumard2006, Lu2013}, M31 \citep{Bender2005, Georgiev2014, Carson2015} and nearby early-type galaxies \citep{Nguyen2017, Nguyen2019}, however, light-weighted ages for the full cluster that are as young as those in our sample are uncommon for typical NSCs. The ages of our sample are more typical for young super star clusters
but the masses ($\sim10^7$\msun) are larger than typical super star clusters by more than an order of magnitude \citep[e.g.,][]{Melo2005, Reines2008}.

The prevalence of galaxies falling within the WISE AGN-selection box is quite low: $\sim$0.2\% in \citet{Hainline2016}. On the other hand, NSCs being found in low-mass galaxies is quite common, implying that the presence of any generic NSC is unlikely to be the cause of the AGN-like WISE colors. However, we emphasize that our sample is extremely unique in just how young the light-weighted ages we calculate for our NSCs are. While the masses and central positions of the clusters are consistent with NSCs, their ages are more in line with super star clusters/young massive clusters. Looking at samples of NSCs in late-type galaxies, both \citet{Kocharov2018} and \citet{Sarzi2005} find light-weighted ages only down to $\mathcal{O}$(100 Myr) and (except for one galaxy) both \citet{Walcher2006} and \citet{Rossa2006} calculate light-weighted ages down to $\sim$30 Myr. Out of the $\sim$40 galaxies presented in these papers, only one galaxy (NGC 2139) has a light-weighted age comparable to those in our sample ($<$10 Myr).

These works also examine the detailed star formation histories (SFHs) of the NSCs in their samples using composite stellar population fits. These SFHs reveal a series of increased periods of star formation followed by periods of quiescence \citep{Rossa2006, Kocharov2018}, meaning that most NSCs contain populations of stars at a variety of ages. When present, populations of stars as young as those that we find in our target galaxies ($<$10 Myr) generally make up $\lesssim$1\% of the mass of the clusters \citep{Walcher2006, Kocharov2018} and, as mentioned above, they find light-weighted ages much older than what we find for our targets.

Only the one galaxy (NGC 2139) in \citet{Walcher2006} has $\sim$7\% of the mass made up by stars with age 3 Myr and they conclude that it is either a young NSC in the process of forming or its bright young population is outshining the underlying, older population; an old population (10 Gyr) with equal mass as the young population would contribute only 1\% of the light. Because of this and since we are calculating ages based on the integrated light of the entire cluster, we cannot distinguish between NSCs that are initially forming or NSCs that have had a recent burst of star-formation. However, either way, finding ages as young as we do for our targets means that a significant burst of star formation would have had to occur within the past 10 Myr, such that this new burst would outshine the rest of the older stars in the cluster.

The high optical extinctions, young ages and large masses of the clusters support the idea that these NSCs are likely dominating the mid-IR emission of the dwarf galaxies and mimicking the {\it WISE} colors of AGNs. However, we cannot definitively rule out the presence of AGNs or quiescent BHs in these objects. Indeed, there are multiple examples of NSCs and central BHs coexisting, such as M31, M32, NGC 3115, and the Milky Way \citep{Neumayer2020}. Specifically in our sample, it is possible that ID 11 has both a young, central star cluster and a central AGN, causing the variability in the mid-IR observed in \citet{Secrest2020}.

Reliably identifying AGNs in local dwarf galaxies is particularly important in the context of understanding the origins of massive BH seeds. Along with other works \citep{Hainline2016, Latimer2021}, we advise caution when using mid-IR color-color selection techniques for dwarf galaxies (at least at the resolution of {\it WISE}). In this work, we present a sample of galaxies hosting remarkably young, dusty and massive star clusters in their nucleus, which negates the need to invoke the presence of an AGN in these galaxies by providing a plausible alternative explanation for the WISE observations; our observational study directly demonstrates that young, massive, dusty NSCs can indeed mimic the mid-IR colors of AGNs. \\


We thank the referee for helpful comments that improved the manuscript. Support for Program number HST-GO-15607.001-A was provided through a grant from the STScI under NASA contract NAS5-26555. AER acknowledges support provided by NASA through EPSCoR grant number 80NSSC20M0231 and the NSF through CAREER award 2235277.


\bibliography{main}

\end{document}